\documentclass[aps, prb, twocolumn, showpacs, preprintnumbers,
amssymb, floatfix]{revtex4}

\usepackage{amsfonts}%
\usepackage{amssymb, amsmath}%
\usepackage{graphicx}
\usepackage{graphicx, , setspace, epstopdf, fancyhdr, mathrsfs, multirow}
\usepackage{subfigure}
\usepackage{dcolumn}
\usepackage{enumerate}
\usepackage{bm}
\usepackage{xcolor}

\newcommand{\colvec}[1]{{\bf{#1}}}

\renewcommand{\d}{\text{d}}

\newcommand{\aref}[1]{Appendix~\ref{#1}}
\newcommand{\Eref}[1]{Equation~\eqref{#1}}

\newcommand{\eref}[1]{Eq.~\eqref{#1}}
\newcommand{\esref}[1]{Eqs.~\eqref{#1}}
\newcommand{\Fref}[1]{Figure~\ref{#1}}
\newcommand{\Fsref}[1]{Figures~\ref{#1}}
\newcommand{\fref}[1]{Fig.~\ref{#1}}
\newcommand{\fsref}[1]{Figs.~\ref{#1}}
\newcommand{\Rref}[1]{Ref.~\onlinecite{#1}}

\newcommand{\sref}[1]{Sec.~\ref{#1}}
\newcommand{\ssref}[1]{Secs.~\ref{#1}}

\newcommand{\tref}[1]{Tab.~\ref{#1}}

\begin{document}

\title[Toroidal dipole excitations in metamolecules
 formed by interacting nanorods]
 {Toroidal dipole excitations in metamolecules
 formed by interacting plasmonic nanorods}
 \author{Derek W. Watson}
 \author{Stewart D. Jenkins}
 \author{Janne Ruostekoski}
 \affiliation{Mathematical Sciences and Centre for Photonic
	Metamaterials, University of Southampton,
	Southampton SO17 1BJ, United Kingdom}
 \author{Vassili A. Fedotov}
 \author{Nikolay I. Zheludev}
 \affiliation{Optoelectronics Research Centre and
 Centre for Photonic Metamaterials
	University of Southampton, Southampton SO17 1BJ, United Kingdom}
	
\begin{abstract}
 We show how the elusive toroidal dipole moment appears as a
 radiative excitation eigenmode in a metamolecule resonator that is
 formed by pairs of plasmonic nanorods. We analyze one such
 nanorod configuration -- a toroidal metamolecule. We find that the radiative
 interactions in the toroidal metamolecule can be qualitatively represented
 by a theoretical model based on an electric point dipole arrangement.
 Both a finite-size rod model and the point dipole approximation demonstrate
 how the toroidal dipole moment is subradiant and difficult
 to excite by incident light.
 By means of breaking the geometric symmetry of the metamolecule,
 the toroidal mode can be excited by linearly polarized
 light and appears as a Fano resonance dip
 in the forward scattered light.
 We provide simple optimization protocols for maximizing the
 toroidal dipole mode excitation.
 This opens up possibilities for simplified control and driving of
 metamaterial arrays consisting of toroidal dipole unit-cell resonators.
 \end{abstract}
 \date{\today}
 \maketitle

 \section{Introduction}\label{sec:Intro}

 The static toroidal dipole, also known as anapole, was first considered by
 Zel'dovich in 1957\cite{Zeldovich}. The toroidal dipole has since been
 acknowledged in nuclear and atomic physics, where it is held
 responsible for parity violations in electroweak
  interactions\cite{PhysRevE.65.046609,Haxton21031997,PhysRevC.56.1641,
  Dmitriev2004243,Flambaum},
 as well as in solid-state physics, where it leads
  to a special type of ordering in multiferroics
 \cite{Schmid,PhysRevB.76.214404,2014FrMat...1....9P,PhysRevB.90.024432}.

 The notion of the tororidal dipole was eventually
 extended to the dynamic case,
 where it generated a whole family of radiating toroidal
 multipoles\cite{Dubovik1990145,Dubovik,PhysRevE.57.6030,PhysRevE.65.046609},
 commonly omitted from standard literature on the
 electrodynamics\cite{Jackson,Morse,LANDAU}.
 While electric and magnetic multipoles are respectively linked to the
 oscillations of charges and transverse currents, the toroidal multipoles
 result from the oscillations of radial currents, which tend to be
 neglected in the long wavelength limit \cite{Morse,Dubovik1990145}.
 The lowest-order toroidal multipole--toroidal dipole--is produced
 by currents circulating on a surface of an imaginary torus along its
 meridians\cite{57065613}.

 The toroidal response in natural materials is weak and usually masked by
 conventional electromagnetic (EM) effects, but with the advent of
 metamaterials and modern nanofabrication technology, materials
 composed of subwavelength particles may be designed and constructed
 to exhibit an enhanced toroidal dipole response.
 Various metamaterial designs have already
 been utilized experimentally to promote a
  toroidal dipole response in the microwave and optical part of the spectrum:
  circular apertures in a metallic screen\cite{FedZheletc},
 asymmetric  split rings\cite{PhysRevB.87.115417},
  split   rings\cite{1402-4896-88-5-055002,Kaelberer10122010},
  and double bars \cite{ZhangX1}.
  In numerical simulations, also other resonator configurations 
 \cite{Huang:12,HuangYaoetc,PhysRevB.89.205112,PhysRevB.87.245429,Dong:12} 
  have  shown notable toroidal dipole responses.

 Thus far, the theoretical understanding of the toroidal dipole
 response in resonator systems has been limited and the conditions
 under which the toroidal moment may be excited on the
 microscopic level have not been well known.
 In this paper we show theoretically how a simple
 structure formed by interacting
 plasmonic nanorods can support a collective excitation
 eigenmode that corresponds to a radiating toroidal moment.
 The toroidal mode is subradiant which
 could be important, e.g., for the applications of the toroidal
 moments in nonlinear optics\cite{Wang199615,CANFIELD}
 and in surface plasmon sensors\cite{Maier}.
 We analyze the light-induced interactions between the closely spaced
 plasmonic rods using a finite-size rod model as well as a model
 where the metamolecule is represented by a simple arrangement
 of point electric dipole emitters. We find that the point dipole
 model provides an accurate description of the radiative properties,
 except at very short inter-rod separations.

 The generally weak coupling of the toroidal dipoles to
 external radiation fields makes it  difficult to excite the
 toroidal mode. The structural symmetry of the mode inhibits
 the coupling to EM field modes that do not possess a
 similar `vortexlike' symmetry. We show how a simple
  linearly polarized incident light beam, however, can drive the
  toroidal dipole excitation when the geometric symmetry of the
  metamolecule is broken. The method is related to the double-resonator
  configuration introduced in \Rref{ZhangX1} and we
  provide simple optimization protocols for maximizing the toroidal
  dipole mode excitation.
  The emergence of the toroidal dipole excitation is shown in the forward
  scattered light as a Fano resonance, indicating a destructive
   interference between the broad-resonance electric dipole
    and narrow-resonance toroidal dipole modes.
  Using linearly polarized beams for the excitation
 of the toroidal dipole mode can be
  especially beneficial in driving and controlling large metamaterial
  arrays of toroidal unit-cell resonators, since the coupling in this
  case is independent of the array or the beam symmetries.

 The structure of this paper is as follows: in \sref{sec:MADyn}, a brief review
 of metamaterial dynamics as formulated in \Rref{PhysRevB.86.085116}
  is presented.
 In \sref{sec:MAinteractions}, the general model is applied to the case of
 cylindrical nanorods. Two models are developed: a
 point dipole model; and a finite-size resonator model.
  In \sref{sec:T1},
 we investigate a toroidal metamolecule comprised of
 plasmonic nanorods. The collective radiative excitation
  eigenmodes, and their resonance linewidths
 and line shifts
 for the toroidal metamolecule are calculated.
  We then show how a toroidal dipole
  response may be excited with linearly polarized light by
  introducing a geometric asymmetry in the metamolecule.
 Some concluding remarks are included in
 \sref{sec:Conclusions}.

 \section{Dynamics of metamaterial systems}\label{sec:MADyn}

 In order to analyze the EM interactions between
 plasmonic rods in a toroidal metamolecule and the
 excitation of toroidal dipole mode by an incident field
 we first briefly introduce the basic formalism of radiatively
 coupled rods. We show how in a simple model the
 incident field and the scattered EM fields from the other rods
 excite normal modes of current oscillations in each rod.
 A more detailed derivation of the formalism,
 also including magnetodielectric systems, can be
 found in \Rref{PhysRevB.86.085116}.

 In the general model of circuit resonator interactions with
 EM fields, we assume that the charge and current sources
  are driven by an incident electric displacement field
 ${\bf D}_{\text{in}}({\bf r},t)$, and magnetic
 induction ${\bf B}_{\text{in}}({\bf r},t)$
  with frequency $\Omega_0$.
  When analyzing the EM fields and resonators,
 we adopt the rotating wave approximation
 where the dynamics is dominated by $\Omega_0$.
 In the rest of the paper, all the EM fields and resonator amplitudes refer
 to the slowly-varying versions of the positive frequency components of the
 corresponding variables, where the rapid oscillations $e^{-i\Omega_0t}$
 due to the dominant laser frequency has been factored out.

 The state of current oscillation in the resonator $j$ may be described by a
 single dynamic variable with units of charge $Q_j(t)$ and
  its rate of change $I_j(t)$, the current.
 The current oscillations within each resonator $j$ behave
 in a manner analogous to an LC circuit with resonance frequency
 $\omega_j$:
 \begin{equation}
 \omega_j = \frac{1}{\sqrt{L_jC_j}}
  \,\textrm{,}
 \label{eq:omega}
 \end{equation}
 where $C_j$ and $L_j$ are an effective self-capacitance
 and self-inductance, respectively.
  In \Rref{PhysRevB.86.085116}, a general theory was formulated
 to derive a coupled set of linear equations for the EM fields and strongly
 coupled resonators. In order to express the coupled equations for the
 EM fields and resonators we introduce
 the slowly varying normal mode oscillator amplitudes $b_j(t)$,
 \begin{equation}
 b_j(t) = \frac{1}{\sqrt{2\omega_j}}
 \left[\frac{Q_j(t)}{\sqrt{C_j}} + i \frac{\phi_j(t)}{\sqrt{L_j}}\right]
  \,\textrm{.}
 \label{eq:b}
 \end{equation}
 Here, the generalized coordinate for the
 current excitation in the resonator $j$ is
 the charge $Q_j(t)$ and $\phi_j(t)$ represents its conjugate momentum.
 In the rotating wave approximation
 the conjugate momentum  is
 linearly-proportional to the
 current\cite{PhysRevB.86.085116}.
 The dynamic variable in \eref{eq:b} can be
 used to describe a general resonator
 with both polarization and magnetization sources.
  The oscillation of $Q_j(t)$ and $I_j(t)$ within each
 resonator are proportional to the  corresponding
  polarization ${\bf P}_j({\bf r},t)$ and magnetization
 ${\bf M}_j({\bf r},t)$ sources\cite{PhysRevB.86.085116}.
 The electric ${\bf E}_{\text{sc},j}({\bf r},t)$
 and magnetic  ${\bf H}_{\text{sc},j}({\bf r},t)$
 fields scattered by emitter $j$,
 are a result of the oscillating polarization and magnetization sources.

\begin{figure}[h!]
  \centering
 \includegraphics[width=0.8\columnwidth]{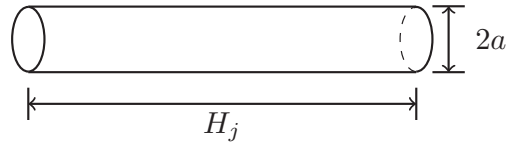}
 \caption{\label{fig:Meta-Atom}Geometry of a single
  nanorod with length $H_j$ and radius $a$.}
 \end{figure}

 We approximate the  resonators as cylinders (nanorods) with
 the radius $a$ and length $H_j$, see \fref{fig:Meta-Atom}.
 In this work we only consider
 a single radius for all nanorods, see \aref{sec:Drude}.
 In \sref{subsec:Driving}
 and~\ref{subsec:Amp} we vary the length of each nanorod
 about a fixed length $H_0$.
 For simplicity, the charge and current oscillations
 within the cylinder are assumed to be linear
 along its axis.
  The magnetization of such a system is
 negligible  ${\bf M}_j({\bf r},t)\simeq 0$.
 The scattered EM fields are thus determined
 by the polarization  ${\bf P}_j({\bf r},t)$ within each nanorod alone,
  resulting in accumulation of charge on the
 ends of the rod.

 The scattered electric field is then given by
 \begin{equation}
 {\bf E}_{\text{sc},j}({\bf r})
 =
 \frac{k^3}{4\pi\epsilon_0}\int\d^3r'\,
 {\bf G}({\bf r} -{\bf r}')\cdotp{\bf P}_j({\bf r}')
  \,\textrm{,}
  \label{eq:Esc}
  \end{equation}
 where $k=\Omega/c$. The radiation kernel
 ${\bf G}({\bf r} - {\bf r}')$ determines
 the electric  field at ${\bf r}$, from the polarization
  source at ${\bf r}'$\cite{Jackson}.
 The total electric fields external to resonator $j$ comprise
 the incident fields and those  scattered
 from all other emitters,
 \begin{equation}
 {\bf E}_{\text{ext},j}({\bf r},t) =
 \frac{1}{\epsilon_0}{\bf D}_{\text{in}}({\bf r},t)
 + \sum_{j\ne j'}{\bf E}_{\text{sc},{j'}}({\bf r},t)
 \label{eq:Eext}
 \end{equation}
 \Eref{eq:Esc} gives the total scattered
 electric field as a function of the
 polarization  density, and is not readily solved for
 ${\bf P}({\bf r},t)$.
 When the relative separation between emitters is less than,
 or of the order of a
 wavelength, a strongly coupled system results.

 We write the polarization density of the
 nanorod as\cite{PhysRevB.86.085116}
 \begin{equation}
 {\bf P}_j({\bf r},t)  = Q_j(t){\bf p}_j({\bf r})
  \,\textrm{,}
 \label{eq:P}
 \end{equation}
 where we assume that the charge profile function ${\bf p}_j({\bf r})$ may be
 considered to be independent of time.
 The driving of the charge oscillations within
 the nanorod  is then provided  by the
 external electric   field ${\bf E}_{\text{ext},j}({\bf r},t)$
  aligned along the direction of the
 source, providing a net electromagnetic
 force (emf)\cite{PhysRevB.86.085116},
 $\mathcal{E}_{\text{ext},j}$,
 \begin{equation}
 \mathcal{E}_{\text{ext},j}(t)=
 \frac{1}{\sqrt{\omega_jL_j}}\int\d^3r\,{\bf p}_j({\bf r})
 \cdotp{\bf E}_{\text{ext},j}({\bf r},t)
  \,\textrm{.}
 \label{eq:emf}\\
 \end{equation}

 For a system which comprises $N$ resonators, the
 collective interactions of the resonators
 with each other and the external field is represented by the linear set
 of equations\cite{PhysRevB.86.085116}
 \begin{equation}
 {\bf \dot b} = \mathcal{C}{\bf b} + {\bf F}_{\text{in}}
  \,\textrm{,}
 \label{eq:equmot}
 \end{equation}
 where ${\bf b}$ is a vector of normal
 oscillator variables and ${\bf F}_{\text{in}}$
 is a vector describing the interaction of
 resonator $j$ with the incident EM field.
 The incident electric  displacement field
 ${\bf D}_\text{in}({\bf r},t)$ with polarization
 vector ${\bf \hat e}_{\text{in}}$, is
 \begin{equation}
 {\bf  D}_\text{in}({\bf r},t) =
 D_\text{in}{\bf \hat e}_{\text{in}}e^{iky}
  \,\textrm{.}
 \label{eq:Ein}
 \end{equation}
 The exact form of $\mathcal{C}$ depends on solving the
 scattered electric field, \eref{eq:Esc},
  for the polarization sources.
 The diagonal elements are a result of the resonator interacting
 with its self-generated EM fields,
 giving rise to the resonance frequency and decay rate
 of a single rod. The off-diagonal elements are
  the result of interactions between different resonators.

 \section{Resonator interactions}\label{sec:MAinteractions}

In the previous section, we briefly introduced the
 model for radiative coupling between different
 emitters that we will apply to the analysis of the
 toroidal metamolecule consisting of plasmonic rods.
 In this section, we focus on two approximations:
 a point emitter model and a finite-size model.
 The finite-size model is an extension of the
 previous treatment\cite{PhysRevB.86.085116}
 and incorporates the corrections due to a finite-size
 polarization distribution that modifies the
  EM interactions between closely spaced rods.

 For the nanorods in \fref{fig:Meta-Atom}, the electric field can be obtained
 using \eref{eq:Esc} with a suitable
 choice of the spatial profile of the charge excitations ${\bf p}_j({\bf r})$.

 \subsection{Point dipole approximation}\label{subsec:PointModel}

 In the point dipole approximation, each
 resonator is modeled as a point electric
 dipole located at the center of mass of the resonator.
 The mode function ${\bf p}_{j}({\bf r})$,
 of the electric dipole is defined as
 \begin{equation}
 {\bf p}_{j}({\bf r})= H_j{\bf \hat d}_j\delta({\bf r} - {\bf r}_j)
  \,\textrm{,}
 \label{eq:pd}
 \end{equation}
 where the proportionality constant $H_j$ has units of
 length and the unit vector ${\bf \hat d}_j$ indicates the
 dipole orientation.  The interaction of the resonator
 with its self-generated EM fields causes radiative damping to occur.
 The rate at which resonators radiate in the dipole approximation
 is $\Gamma_{\text{E},j}$~\cite{PhysRevB.86.085116}, where
 \begin{equation}
 \Gamma_{\text{E},j}=\frac{C_jH_j^2\omega_j^4}{6\pi\epsilon_0c^3}
  \,\textrm{.}
 \label{eq:GammaE1}
 \end{equation}
 We account for non-radiative losses by adding the phenomenological decay
  rate $\Gamma_{\text{O},j}$. The total decay
 rate is then the sum of the radiative
  emission rate and ohmic losses
    \begin{equation}
  \Gamma_j = \Gamma_{\text{E},j} + \Gamma_{\text{O},j}
  \,\textrm{.}
 \label{eq:Gamma_total}
 \end{equation}
 The point dipole approximation, with the inclusion of
 magnetic dipoles, has previously
 been successfully applied to the studies of collective
  effects in planar resonator arrays, e.g.,
 the transmission properties\cite{JenkinsLineWidthNJP,PhysRevLett.111.147401}
 and the development of an electron-beam-driven light source from the
 collective response\cite{PhysRevLett.109.217401}.

 \subsection{Finite-size resonator model}\label{subsec:finite}

 Next, we consider the scattering of EM fields from a collection
 of finite-size nanorods, described earlier
 (see \fref{fig:Meta-Atom}), by integrating over a
  constant density of atomic dipoles over the nanorod's volume.

  \begin{figure}[h!]
   \centering
   \includegraphics[width=0.75\columnwidth]{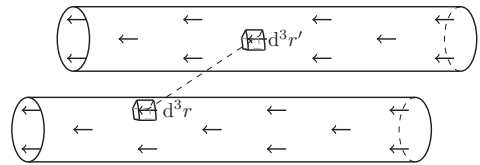}
   \caption{\label{Drg:2RodMC}Schematic of the interaction
    between two nanorods, which
   comprise a uniform distribution of electric
   dipoles (arrows) aligned along the axis of the cylinder.
   }
   \end{figure}

 The scattered electric field from nanorod $j$
  due to polarization sources alone is,
  using \esref{eq:Esc} and~\eqref{eq:P},
 \begin{equation}
 { \bf E}_{\text{sc},j}({\bf r},t) =
 \frac{Q_jk^3}{4\pi\epsilon_0}
 \int\d^3r'\,{\bf G}({\bf r} - {\bf r}')\cdotp {\bf p}_{j}({\bf r}')
  \,\textrm{.}
 \label{eq:EscP}
 \end{equation}
 In the limit $a,H\ll \lambda$,
 the analytic solution for the scattered EM fields from
  the nanorod in the far field zone yields the result
  of an oscillating point dipole,
 with the corresponding radiative linewidth
 of a point dipole of \eref{eq:GammaE1}.
 In \aref{sec:Drude}, we will describe how we estimate the radiative
 and ohmic decay rates for gold nanorods.

   Whilst the far field radiation of a small nanorod
  behaves like that of a point dipole,
    the near and intermediate fields are much more complex,
 with the spatial dependence
 contributions varying as $1/r^3$ and $1/r^2$, respectively.
  The full field solution to \eref{eq:EscP}
  must be integrated numerically, also taking into
  account the finite thickness of the rods.
  The driving of nanorod $j$ is provided by an external
  emf \eref{eq:emf}, which can be broken into contributions from the
  incident and scattered EM fields
  \begin{equation}
  \mathcal{E}_{\text{ext},j} = \mathcal{E}_{\text{in},j}
 +
  \sum_{j\ne j'}\mathcal{E}_{j,j'}^\text{sc}
   \,\textrm{.}
 \end{equation}
 The incident emf $\mathcal{E}_{\text{in},j}$ follows from \eref{eq:emf}
 using the incident EM field  \eref{eq:Ein}.
 The emf driving the rod $j$ by the scattered fields from the rod $j'$,
 $\mathcal{E}_{j,j'}^\text{sc}$,  is obtained
 from \esref{eq:emf} and~\eqref{eq:EscP}
    \begin{equation}
 \mathcal{E}_{j,j'}^\text{sc}=
 \frac{Q_{j'}k^3}{4\pi\epsilon_0\sqrt{\omega_jL_j}}
  \int\d^3r\d^3r'\,{\bf p}_j({\bf r})\cdotp{\bf G}({\bf r} - {\bf r}')
  \cdotp{\bf p}_{j'}({\bf r}')
   \,\textrm{.}
  \label{eq:emf_jj'}
  \end{equation}
  \Eref{eq:emf_jj'} consists of the integral over the volume
 of the nanorod  located at ${\bf r}$ and the volume of the
 nanorod located at ${\bf r}'$,
 see, e.g., \fref{Drg:2RodMC}.

 \section{Toroidal metamolecule}\label{sec:T1}

 In this section we will use the theoretical
 models introduced in \ssref{sec:MADyn}
 and~\ref{sec:MAinteractions} to analyze the
 radiative interactions between plasmonic nanorods
 that form the toroidal metamolecule.
 We first show how a toroidal metamolecule
 composed of nanorods may exhibit a
 toroidal dipole response and calculate the
 eigenmodes of radiative excitations in the
 toroidal metamolecule. We then show how
 the toroidal metamolecule can be excited by
 incident EM fields when the symmetry of the metamolecule is broken.

 Toroidal dipoles are formed by oscillating poloidal currents.
 The weakness of the toroidal dipole in comparison to its
 electric and magnetic counterparts in natural materials mean it
 is often neglected in classical physics\cite{Jackson,Morse,LANDAU}.

 We introduce a toroidal metamolecule that generally comprises $N$ nanorods
 distributed over two layers.
 Each layer has $N_\theta=N/2$ emitters orientated radially
 outwards from the central
  axis of the metamolecule and equally spaced in the
 azimuthal direction. The second layer
 of emitters, identical to the first, is positioned a
 distance $2y$ above the first layer. The orientation
 of both layers is such that there are $N_\theta$ pairs of parallel emitters.
 The symmetry of the metamolecule is
 therefore $C_{{N\over 2}h}$, a combination
 of $N/2$ rotations about the $N/2$-fold
 symmetry axis $C_{{N\over 2}}$ and
 the reflection in a horizontal plane $\sigma_h$ (a plane perpendicular
 to the principal axis of rotation).

 In Cartesian coordinates, the orientation vectors of the nanorods are
 ${\bf \hat d}_j={\bf \hat x}\cos\theta_j + {\bf \hat z}\sin\theta_j$.
 The density of the metamolecule is determined by
 $\Delta\theta = \theta_{j+1}-\theta_j$.
 As $\Delta\theta\rightarrow 0$, the metamolecule approaches  a torus.

 In the present section, for simplicity, we analyze the radiative
 properties of a toroidal metamolecule consisting of $N=8$ nanorods,
 where $\Delta\theta = \pi/2$. Although
 an eigenmode of such a metamolecule only approximately describes a
 toroidal dipole, the characteristic properties of the resonator interactions
 and the toroidal dipole excitation by an incident field are already evident.
 As the density of the structure increases,
 the analysis can be easily extrapolated to
 account for the increased number of resonators as illustrated in the toroidal
 mode excitation shown in \fref{Drg:Asymmetry}(a-b).

 We first study a symmetric toroidal metamolecule
 (where all the rods are of equal length) and then break
  the geometric symmetry of the metamolecule in order to
 excite the toroidal mode using a simple light beam.
 In the symmetric metamolecule, we identify the associated
  collective modes of current oscillation  and compare the resonance
 linewidths and line shifts obtained both in the point dipole
 approximation and in the finite-size resonator
 models. Finally, we determine how the toroidal
 dipole response of a toroidal metamolecule with some inherent asymmetry
 may be driven by linearly polarized light.

 \subsection{Eigenmodes of a symmetric toroidal metamolecule}
 \label{sec:symm-toro-metam}

\begin{figure}[h!]
 \centering
 \includegraphics[width=0.75\columnwidth]{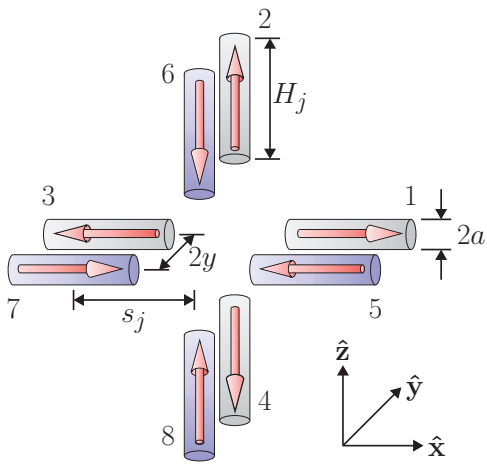}
 \caption{\label{Drg:8RodCartoon}(Color online)  Toroidal
 dipole mode of an eight
 symmetric rod metamolecule. The shading of the rods
 indicates those rods in a shared plane,
 the arrows indicate the phase of current oscillation.
 The radial position of the center of mass for
 an individual rod is $s_j$.
 The separation between parallel layers is $2y$.}
 \end{figure}

 A schematic illustration showing the arrangement and labeling system for the
 nanorods is shown in \fref{Drg:8RodCartoon}.
 In  Cartesian coordinates the locations of the
 point dipole and the center of mass of
 the finite-size nanorods are,
 \begin{eqnarray}\label{eq:locations}
 {\bf r}_{1,5} &=&
 \begin{bmatrix} s_{1,5}\\\pm y\\0
 \end{bmatrix},
 \quad %
 {\bf r}_{2,6} =
 \begin{bmatrix} 0\\\pm y\\s_{2,6}
 \end{bmatrix},\nonumber\\ %
 &&{\bf r}_{3,7} =
 \begin{bmatrix} -s_{3,7}\\\pm y\\0
 \end{bmatrix},
 \quad%
 {\bf r}_{4,8} =
 \begin{bmatrix} 0\\\pm y\\-s_{4,8}
 \end{bmatrix}
  \,\textrm{.}
 \end{eqnarray}
 In a symmetric system then $s_j = s$ (for all $j$).
 Because the nanorods are symmetric, the radiative emission rate,
 ohmic loss rate and total decay rate
 of each rod are identical, i.e., $\Gamma_{\text{E},j}=
 \Gamma_{\text{E}}\,,\Gamma_{\text{O},j}
 =
 \Gamma_{\text{O}}$,
  and
  $\Gamma_j = \Gamma$. In \aref{sec:Drude} we calculate the resonance
  frequency $\omega_0$ and relative decay rate
   $\Gamma_{\text{O}}/\Gamma_{\text{E}}$
  as a function of the rod length for a finite-size rod using formulas for
  resonant light scattering from metal particles developed
  in \Rref{kuwate}, where ohmic losses are incorporated
 in the analysis by the Drude model. In order to
  simplify the comparisons between the point dipole
  approximation and the finite-size model, 
 we use in the both models the same values for the
 decay rates.
  We choose the parameters for the gold nanorods of
 the symmetric metamolecule
 so that the length $H_0=1.5\lambda_p\simeq 209$nm,
 where $\lambda_p$ denotes the plasma wavelength
 for  gold  (see Appendix A). This yields $H_0\simeq0.243\lambda_0$ 
 and $a\simeq 0.0324\lambda_0$, where
  $\lambda_0=2\pi c/\omega_0\simeq859$nm 
denotes the resonance wavelength of the nanorod.
 We use these dimensions as those of
 a reference nanorod throughout the paper. The corresponding decay rates 
 $\Gamma_{\rm E}\simeq0.83\Gamma$
 and $\Gamma_{\rm O}\simeq0.17\Gamma$, see Fig. 12 and Appendix A.
 The choice of the parameters ensures that the
 decay rates are only weakly sensitive to
 the small changes of the rod length.

 The incident light, tuned to the resonance frequency of the nanorods
 $\Omega_0=\omega_0$, drives the charge oscillations within the
 nanorods. Each nanorod scatters light due to its polarization density
 that results from the
 charge oscillations. The light can multiply scatter between
 different resonators. Strong
 multiple scattering results in collective excitation modes
  in the system of nanorods. The
 collective eigenmodes can then exhibit different resonance
 frequencies and linewidths and line
  shifts\cite{PhysRevB.86.085116,JenkinsLineWidthNJP,PhysRevLett.111.147401}.
 We calculate the collective eigenmodes
 of the interacting rod configuration
 by analyzing the light-induced interactions in the equations of
 motion for the resonator excitations, \eref{eq:equmot}.
 The collective modes of current oscillation within the system are described by
 the eigenvectors ${\bf v}_n$ of the interaction
 matrix $\mathcal{C}$. The corresponding eigenvalues
 $\xi_n$ have real and imaginary parts corresponding to the decay rate and
  resonance frequency shift of the mode,
 \begin{equation}
 \xi_n = - \frac{\gamma_n}{2} - i(\Omega_n - \Omega_0)
 \,\textrm{.}
 \label{eq:lambda_n}
 \end{equation}
 The number of individual resonators
 determines the number of collective modes.
 The different modes may have superradiant or
 subradiant characteristics. The former occurs when the emitted radiation is
 enhanced by the interactions of the resonators
 ($\gamma_n> \Gamma$), the latter occurs when the
 radiation is suppressed 
 ($\gamma_n < \Gamma$).
 \begin{figure}[h!]
   \centering
 \includegraphics{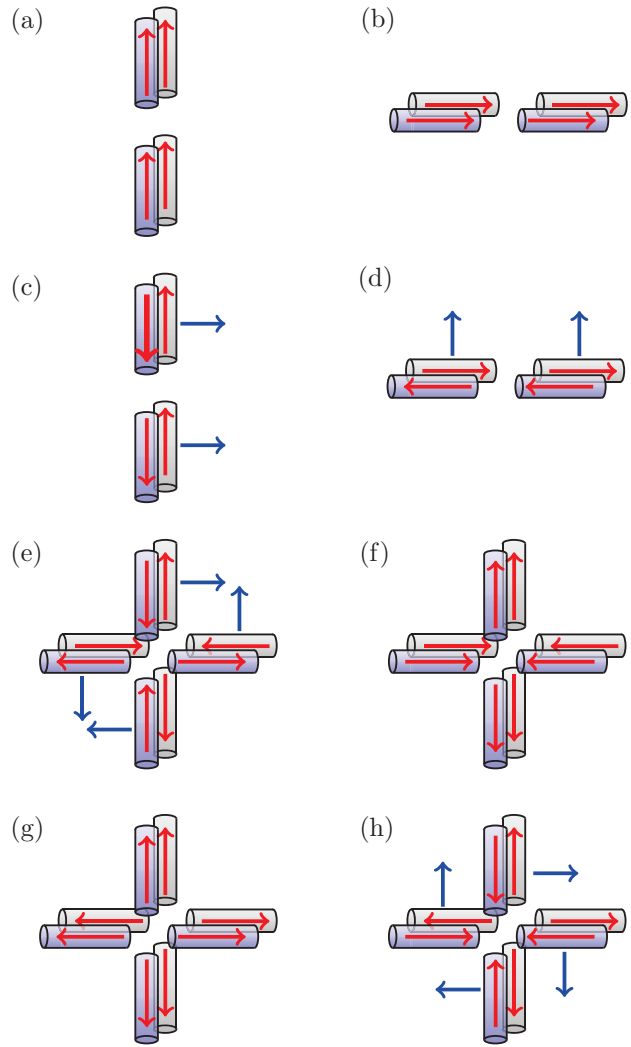}
 \caption[]{\label{fig:Eigenmodes}(Color online)  Representation
 of the eigenmodes of a
 symmetric eight-rod metamolecule.
 The red arrows represent the electric dipole
 moments and the blue  arrows effective magnetic dipole moments.
 The modes are classified as:~(a) vertical
 and~(b) horizontal electric dipole (E1) modes; (c) vertical and~(d)
 horizontal magnetic dipole (B1) modes; (e)  a magnetic
 quadrupole (B2) mode; (f) an electric
 quadrupole (E2) mode; (g) a symmetric (sy) mode; and~(h)
 the toroidal dipole (t) mode. For shading and axis properties,
 see \fref{Drg:8RodCartoon}.
 }
 \end{figure}
 The classification of the different eigenmodes of the toroidal metamolecule
 comprising of symmetric nanorods is shown in \fref{fig:Eigenmodes}.
 In \tref{tab:C4h},
 we list the character table
  for the symmetry group $C_{nh}$
   (for $n=4$)\cite{character_table,Note}, where
 the eigenmodes are related to the
 corresponding Mulliken symbols of the point group.

\begin{table}[h!]
\caption{\label{tab:C4h} Character table of $C_{nh}$, for $n=4$. The
 terms in parentheses are our physical multipole designation,
 see \fref{fig:Eigenmodes}, for the equivalent Mulliken symbols.
 }
\begin{ruledtabular}
\begin{tabular}{lrrrrrrrr}
$C_{4h}$ &
 $E\footnote{Identity}$
 & $C_{4}$ & $C_{2}$ & $C_{4}^{3}$ &
 $i{\footnote{Inversion}}$
 &
 $S_{4}^{3\footnotesize{\,\, c}}$
 &  $\sigma_{h}$
 &
 $S_{4}\footnote{Improper rotation $S_n=C_n\sigma_h$}$ \\
\hline
A (sy) & $1$ & $ 1$ &  $1$&  $1$ & $1$ &  $1$ & $1$ & $ 1$\\
B (E2)& $1$ & $ -1$ &  $1$&  $-1$ & $1$ &  $-1$ & $1$ &$ -1$\\
E (E1)& $2$ &$ 0$ &  $-2$&  $0$ & $2$ &  $0$ & $-2$ &$ 0$\\
A$'$ (t) & $1$ & $ 1$ &  $1$&  $1$ & $-1$ &  $-1$ & $-1$ &$ -1$\\
B$'$ (B2) & $1$ &$ -1$ &  $1$&  $-1$ & $-1$ &  $1$ & $-1$ &$ 1$\\
E$'$ (B1) & $2$ &$ 0$ &  $-2$&  $0$ & $-2$ &  $0$ & $2$ &$ 0$\\
\end{tabular}
\end{ruledtabular}
\end{table}
 The two modes depicted in \fsref{fig:Eigenmodes}(a)
 and~\ref{fig:Eigenmodes}(b)
 correspond to vertical and horizontal electric dipole (E1) modes, respectively.
 In the E1 modes, the responsive nanorods oscillate in phase and there is
 an effective electric dipole.  Each of the E1 modes is a rotation of the other,
 consequently they experience the same resonance
 frequency shift and decay rate.
 In a similar manner to the E1 modes, \fsref{fig:Eigenmodes}(c)
 and~\ref{fig:Eigenmodes}(d)  depict vertical and
 horizontal magnetic dipole modes (B1),
 respectively. In the B1 modes, pairs of responsive nanorods oscillate out of
 phase, and the metamolecule forms an effective magnetic dipole.
 The B1 modes are also rotations of each other, thus experience the same
 resonance frequency shift and decay rate.

 The remaining modes are all independent. \Fref{fig:Eigenmodes}(e)
 corresponds to a magnetic quadrupole (B2) mode. Each parallel pair of
 emitters oscillates out-of-phase, and four independent effective magnetic
 dipoles form. \Fref{fig:Eigenmodes}(f)
 depicts an electric quadrupole (E2)
 mode. In this mode, parallel pairs of emitters
 oscillate in phase with each other,
 but out of phase with their opposite parallel pair. Four independent effective
 dipoles combine to form an effective electric
 quadrupole.  \Fref{fig:Eigenmodes}(g)
 is the symmetric mode, where all nanorods oscillate in phase. The toroidal
 dipole mode is shown in \fsref{fig:Eigenmodes}(h)
 and~\ref{Drg:8RodCartoon}. In this mode, parallel
 pairs of emitters oscillate out of phase in such
 a way that the effective magnetic
 dipoles form a circular loop.
 The orientations of the nanorods lead to the orientation vectors
 for their corresponding electric   dipole moments: ${\bf \hat d}_{1,7} =
 {\bf \hat x}$, ${\bf \hat d}_{2,8} = {\bf \hat z}$, ${\bf \hat d}_{3,5} =
  -{\bf \hat x}$, and ${\bf \hat d}_{4,6} =  -{\bf \hat z}$.

 \begin{figure}[h!]
   \centering
 \includegraphics[width=0.9\columnwidth]{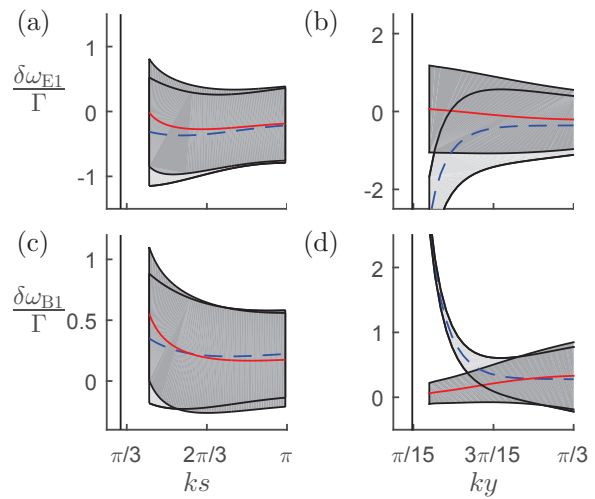}
 \caption{\label{fig:8Rod_E1B1}(Color online)
 The radiative resonance linewidths and line shifts
 for the collective electric dipole (E1) and magnetic
 dipole (B1) excitation eigenmodes, as a
 function of the metamolecule parameters $s$ and $y$.
 We show the line shift in the point dipole model (blue dashed line) and
 finite-size model (red solid line), the linewidth in
  the point dipole model (light shading
 about the blue dashed line), and the linewidth in
 the finite-size model (dark shading about the
 red solid line).
 In~(a), the E1 mode, and in~(c), the B1 mode, are
 shown as functions of the radial position $s$,
 with layer position $y=\lambda_0/6$.
 In~(b), the E1 mode, and
 in~(d), the B1 mode, are shown as
 functions of the layer position $y$, with radial position $s=\lambda_0/4$.
 The finite-size rods have lengths $H_0=0.243\lambda_0$ and
 radii $a=0.0324\lambda_0$. The radiative losses of
 each nanorod are $\Gamma_{\text{E}}=0.83\Gamma$, the ohmic losses are
 $\Gamma_{\text{O}}=0.17\Gamma$.
 }
 \end{figure}

 \begin{figure}[h!]
   \centering
 \includegraphics[width=0.9\columnwidth]{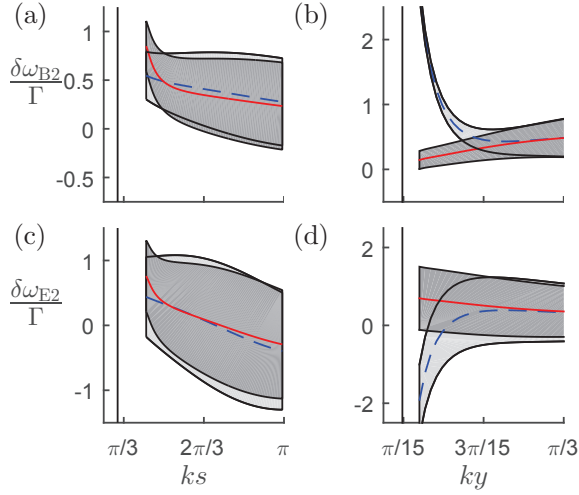}
 \caption{\label{fig:8Rod_B2E2}(Color online)
 The radiative resonance linewidths and line shifts
 for the collective magnetic quadrupole (B2) and electric quadrupole
 (E2) excitation eigenmodes, as a
 function of the metamolecule parameters $s$ and $y$.
 In~(a),
 the B2 mode, and in~(c),  the E2 mode,
 are shown as functions of the radial position $s$.
 In~(b), the B2, and in~(d),
 the E2 mode, are shown as functions of the layer position $y$.
 For the rod parameters and plot
 descriptions see \fref{fig:8Rod_E1B1} caption.
 }
 \end{figure}

\begin{figure}[h!]
   \centering
 \includegraphics[width=0.9\columnwidth]{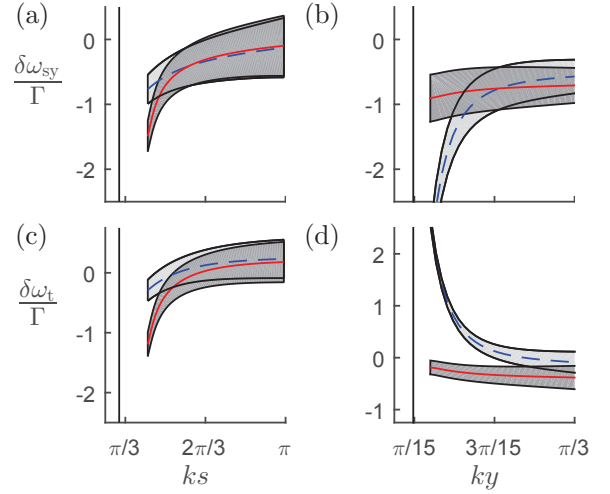}
 \caption{\label{fig:8Rod_syT1}(Color online)
 The radiative resonance linewidths and line shifts
 for the collective symmetric and toroidal dipole excitation eigenmodes, as a
 function of the metamolecule parameters $s$ and $y$.
 In~(a)
 the symmetric mode, and~(c) the toroidal dipole
 mode are shown as a function of the radial
 position $s$.
 In~(b) the symmetric mode and~(d) the toroidal dipole mode are shown as
 a function of the layer position $y$.
 For the rod parameters and plot descriptions
 see \fref{fig:8Rod_E1B1} caption.
 }
 \end{figure}

 We calculate the collective eigenmodes of the dynamic system described by
 \eref{eq:equmot}, using the point dipole approximation and the
  finite-size resonator model discussed in \ssref{subsec:PointModel}
   and~\ref{subsec:finite}.
 In \fsref{fig:8Rod_E1B1}-\ref{fig:8Rod_syT1}, the  line shifts and
 the corresponding resonance linewidths are shown, for the collective
  modes, as  functions of the
 metamolecule varying parameters $s$  and $y$ (in the former the
 layer spacing is fixed at $y = \lambda_0/6$ and in the latter the
 radial spacing is fixed at $s = \lambda_0/4$).
 These modes are depicted in \fsref{fig:Eigenmodes}(a)--\ref{fig:Eigenmodes}(g),
 respectively.

 We find that the point dipole model qualitatively agrees with the finite-size
 model. The agreement becomes very good
 for larger rod separations. The toroidal
 dipole mode and the symmetric mode are subradiant
 for all parameter values we considered.
 This is also true for the magnetic quadrupole mode,
 B2, except for some specific values
 of the rod positions. The toroidal dipole mode is always
 the most subradiant mode, indicating
 a very weak coupling to external light fields.
 Magnetic dipole and electric quadrupole modes can generally exhibit
 both superradiant or subradiant characteristics depending on
 the precise details of the metamolecule's construction, while the
 electric dipole modes are almost always superradiant.

 Specifically, for small radial separation $s = \lambda_0/6$ (and
 the layer separation is fixed at $y=\lambda_0/6$),
 the superradiant modes, E1 and B1
  [\fsref{fig:8Rod_E1B1}(b) and~\ref{fig:8Rod_B2E2}(a)],
 have decay rates $\gamma_\text{E1}\approx 1.8\Gamma$
 and $\gamma_\text{B1}\approx 1.15\Gamma$.
 The toroidal dipole mode has $\gamma_\text{t} \approx 0.3\Gamma$.
 When $s\approx \lambda_0/4$,
 the  E2 mode also becomes
 superradiant  ($\gamma_{\text{E2}}\approx 1.25\Gamma$), while
 the decay rates of the E1 and B1 modes reduce to
 $\gamma_\text{E1}\approx   1.5\Gamma$
 and $\gamma_\text{B1}\approx \Gamma$, respectively. Even for
 the larger separation the toroidal dipole mode is still
 strongly subradiant ($\gamma_\text{t}\approx 0.4\Gamma$
 at $s\approx  \lambda_0/4$ and
 $\gamma_\text{t}\approx 0.6\Gamma$ at $s\approx \lambda_0/2$).
 For large  radial positions,
 $s\approx \lambda_0/2$, only the E1 and E2 modes are superradiant
 ($\gamma_\text{E1}\approx 1.2\Gamma$ and
  $\gamma_\text{E2}\approx 1.65\Gamma$, respectively)
 and the B2 mode  becomes
  subradiant ($\gamma_\text{B1}\approx 0.75\Gamma$).

 When we reduce the layer spacing to $y<\lambda_0/16$, with the
  radial separation fixed at $s = \lambda_0/4$, several of the
 modes become subradiant,
 except the E1 and E2 modes ($\gamma_{\text{E1}}\approx 1.6\Gamma$
  and $\gamma_{\text{E2}}
  \approx 2.2\Gamma$ at $y\approx \lambda_0/16$).
 The toroidal dipole decay rate at $y\approx \lambda_0/16$ is
  reduced to $\gamma_\text{t}\approx 0.25\Gamma$
  from $\gamma_\text{t}\approx 0.45\Gamma$
 at $y \approx \lambda_0/6$.

 \Fsref{fig:8Rod_E1B1}--\ref{fig:8Rod_syT1}
 also display the resonance line shifts of the modes,
  $\delta\omega_n=-(\Omega_n - \Omega_0)$.
 As the radial
 separation becomes large, these asymptotically
 approach a  constant. This is due to the large
 separations resulting in the relatively close
  parallel pairs of nanorods interacting independently as dipoles.
  The B2 mode has the biggest shift
  with $\delta\omega_\text{B2}/\Gamma\approx 0.3$.
  The line shift of the toroidal dipole and E1 modes are
 small $\delta\omega_\text{t}/\Gamma\approx 0.1$
  and $\delta\omega_\text{E1}/\Gamma\approx 0.05$.
  As the separation becomes small,
  $s<\lambda_0/4$, the  line shifts of the finite-size
  model begins to deviate from those calculated in the
  point dipole approximation. At this range the total
  length of the metamolecule is $0.75\lambda_0$ and  finite lengths of the
  nanorods become increasingly important to their interactions.

 When the layer spacing parameter $y$ is varied there is a more pronounced
 deviation in line shifts for small $y$. The line shift of the point dipole model
  begins to deviate when $y\approx 2\lambda_0/15$. Here the total width of
  the metamolecule is $\lambda_0/6$, and the rods'
 finite radii begin to affect their interactions.
  In the region shown for $y$ in \fsref{fig:8Rod_E1B1}--\ref{fig:8Rod_syT1},
  the line shifts of the different modes do not approach  constant values.

 \subsection{Driving the toroidal dipole response}\label{subsec:Driving}

 A natural method of driving the toroidal dipole
 response in a toroidal metamolecule
 is to use radially polarized light. In the paraxial approximation,
  an incident displacement field
 ${\bf D}_\text{in}({\bf r},t) =
 D_\text{in}(\rho,y,t)(\cos\phi{\bf \hat z} + \sin\phi{\bf \hat x})$,
 can be estimated in terms of complex
 vectors ${\bf \hat e}_{+}$ and ${\bf \hat e}_{-}$,
 where ${\bf  \hat e}_{\pm} =
 \mp({\bf \hat z} \pm i{\bf \hat x})/\sqrt{2}$, by
 \begin{equation}
  {\bf D}_\text{in}({\bf r},t) =\frac{D_\text{in}(\rho,y,t)}{\sqrt{2}}
 \left[e^{i\phi}{\bf \hat e}_{-}  - e^{-i\phi}{\bf \hat e}_{+}\right]
  \,\textrm{.}
 \label{eq:E_inrad}
 \end{equation}
 \Eref{eq:E_inrad} is a superposition of Laguerre-Gaussian
 beams with one unit of angular momentum
 which can couple directly to the toroidal dipole mode.

 Employing radially polarized light, the toroidal dipole
 response may be driven in a symmetric toroidal metamolecule.
 When linearly polarized light is shone on the symmetric toroidal
 metamolecule the toroidal dipole response is suppressed;
 the dominant responses driven are the E1 responses.

 Here, we analyze in detail how a toroidal dipole mode can
 also be excited using linearly polarized light and provide
 a simple protocol how to optimize
 the toroidal dipole excitation.
 Linear polarization has an advantage that it is readily available
 in an experiment and can easily be employed to drive
 toroidal dipole modes in large arrays of metamolecules, independently of the
 symmetry of the array or the beam.

 Rather than spatially varying the light field to alter the excitation of
   individual rods (as done in the case of radial polarization) we alter
   the responses of individual rods to linearly
   polarized light by tailoring the length of the rods.
   Introducing the asymmetry in the rod lengths,
   according to \fref{fig:GammaO_Omega0_RelativeGamma_E1}(b),
 shifts the resonance
   frequencies and introduces a geometric asymmetry in the metamolecule.
   A similar principle was phenomenologically
 introduced in \Rref{ZhangX1} where
 asymmetric pairs of nanorods were experimentally employed to produce a
 toroidal dipole response.

   To see how the rods should be altered, we consider an incident
   linearly polarized light wave tuned to the
 resonance frequency $\Omega_0=\omega_0$ of
   our reference nanorod.
   The length of each rod $j$ is then changed
 by $\delta H_j$, whilst the radius is fixed.
   For sufficiently small $\delta H_j$, the alteration
   shifts  the rod's resonance frequency by $\delta \omega_j$
    in proportion to $\delta H_j$, as we demonstrate
    using the Drude model in \aref{sec:Drude},
    see \fref{fig:GammaO_Omega0_RelativeGamma_E1}.
   In this section, we will derive the pattern of rod length
   asymmetries required for linearly polarized light to excite a toroidal dipole in
   the limit that the incident field is far detuned from resonance with
   any individual nanorod, i.e., $\delta\omega_j \gg \Gamma_j$.  In this
   limit, interactions between nanorods can be neglected.
  We demonstrate how this scheme functions with smaller asymmetries in
  the presence of interactions in  \aref{sec:asymm-coupl-coll}.

\begin{figure}[h!]
   \centering
 \includegraphics[width=0.9\columnwidth]{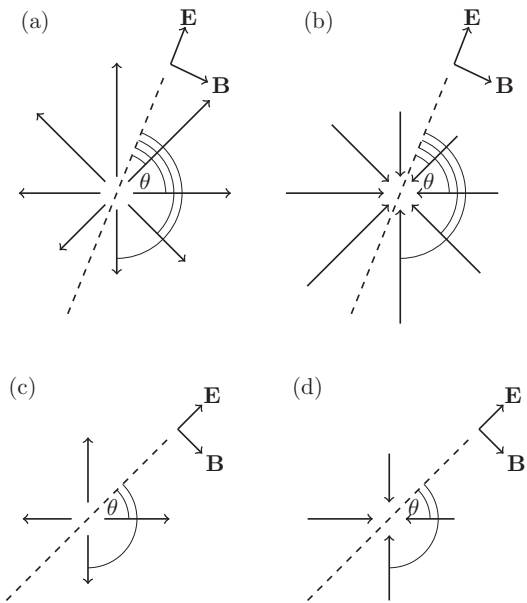}
 \caption{\label{Drg:Asymmetry}The excitation of the toroidal dipole mode
 by linearly polarized light. The length of
 the arrows indicate the rod lengths which,
 together with the angle $\theta$ each rod makes with the polarization of the
 incident light, ensures each rod is equally excited. The arrow
  direction indicates the state of the current oscillation within each nanorod.
  In~(a), the top layer and~(b), the bottom layer of the more general
  $N=16$ case is shown. In~(c), the top layer and~(d), the bottom layer of
  the $N=8$ case is shown. The $N=8$ case is
 considered in the numerical simulation.
 }
 \end{figure}

 We assume that the two layers are separated
 by a distance much less than a wavelength,
 hence the phase difference of the incident field  between layers is negligible.
 In order to couple the field to all nanorods,
 we choose the polarization of the incident
 field to be such that the it bisects the angle created by
 two adjacent nanorods in the same plane, as depicted in
 \fref{Drg:Asymmetry}.
 For a symmetric metamolecule, a field propagating into the plane of
 the metamolecule induces an emf [described by \eref{eq:emf}] driving each
 nanorod $j$ with an amplitude
 \begin{equation}
   F_{\text{sym},j} = F_0\cos\theta_je^{iky_j }
    \,\textrm{,}
   \label{eq:f_in}
 \end{equation}
 where $F_0$  is the driving amplitude of a rod oriented parallel to the
 incident field polarization, and $y_j$ is the position coordinate of
 rod $j$ along the incident field's propagation direction.
 The strength of interaction between the driving field and
 a nanorod varies with the angle $\theta_j$ between the nanorod and
 polarization of the incident light.
 Because the emf induced by the incident field along a rod is
 proportional to its length, the asymmetry in rod lengths perturbs the
 driving strength of each rod $j$ in proportion to $\delta H_j$, so
 that the rod driving is
 \begin{equation}
   \label{eq:F_tot_j}
   F_{j} = (F_0 + \delta F_j)
   \cos\theta_j
   \, \textrm{,}
 \end{equation}
 where $\delta F_j\propto \delta H_j$ is the change in
 driving amplitude rod $j$ would experience if it were parallel to the
 incident field.

 Under these circumstances, when
 $\delta\omega_j \gg \Gamma_j$, interactions between resonators
 can also be ignored (i.e., $\delta\omega_j \gg \mathcal{C}_{jj'}$ for
 $j \ne j'$) in the dynamics of nanorod $j$, and
 \begin{equation}
   \label{eq:single_ma_eq_of_m}
  \dot{b}_j \approx i \delta\omega_j b_j + (F_0+\delta F_j) \cos\theta_j
   \,\textrm{.}
 \end{equation}
 Thus, to lowest order in $\delta H_j$, nanorod $j$ has the
 steady-state response to the incident field
 \begin{equation}
   \label{eq:b_ss}
   b_j \simeq  i \frac{\cos\theta_j}{\delta\omega_j}F_0
   \,\textrm{.}
 \end{equation}
 Therefore, in the noninteracting limit, one could engineer the
 response of a metamolecule simply by adjusting the resonance
 frequencies of its individual rods.

 In a toroidal dipole excitation, all resonators
 in the layer $+y$ oscillate radially outward
 (inward) in phase with each other,
 whilst those resonators in the layer $-y$ oscillate
 radially inward (outward) in phase.
 Such an excitation corresponds to the toroidal dipole eigenmode of
 the completely symmetric metamolecule.
 This is indicated by the arrows in
 \fref{Drg:8RodCartoon}.
 To obtain this excitation profile, \eref{eq:b_ss} suggests that
 the rod lengths should be modified such that
 \begin{equation}
   \delta H_{j}=\left\{
     \begin{array}
       [c]{ll}%
       \delta H_{0}\cos\theta_{j} & \textrm{for }  j=1,\ldots,\frac
                                        {N}{2}  \\
      -\delta H_{0}\cos\theta_{j} & \textrm{for }
                                        j=\frac{N}{2}+1,\ldots, N
 \end{array}
 \right.
 \,\textrm{,}
 \label{eq:general_rod_length_profile}
 \end{equation}
 where $\delta H_0$ is a reference change in rod length, and we have
 assumed that $\delta H_0$ is sufficiently small that $\delta \omega_j
 \propto \delta H_j$.
 Although this profile of rod lengths was arrived at in a regime where
 interactions are neglected, we show that a similar distribution of
 lengths can also be effective in producing a toroidal dipole from a
 linearly polarized light wave driving in \aref{sec:asymm-coupl-coll}.
 We illustrate the dependence of a nanorod's length, and hence
 resonance frequency, on its position within the metamolecule for the
 cases of $N=16$ and $N=8$ nanorods in
 \fref{Drg:Asymmetry}.

 For a toroidal metamolecule, which comprises  eight nanorods,
 only two distinct lengths (resonance frequencies) are required in
 order to produce a toroidal dipole excitation from linearly
 polarized light.
 This equates to a difference in rod length $\mp\delta H$
 about a mean rod length $H_0$. We define $H_{l,s} = H_0 \pm \delta H$
 with $H_{1,2,7,8}=H_l$ and $H_{3,4,5,6}=H_s$, such that each
 parallel pair of rods is composed of a long and short rod, see
 \fsref{Drg:Asymmetry}(c) and~\ref{Drg:Asymmetry}(d), whose
 center's of mass are located at $s_j = s$.
 The polarization vector of the incident field, \eref{eq:Ein},
 that will excite the toroidal dipole mode is
 ${\bf \hat e}_{\text{in}} = ({\bf \hat x} + {\bf \hat z})/\sqrt{2}$.
 The eigenmodes
  and line shifts and widths for a symmetric
 toroidal metamolecule comprising  rods
   with length $H_0$ were discussed in
  \sref{sec:symm-toro-metam}. When the radial position
  of the reference rod is fixed, i.e., $s_l=\lambda_0/3$,
  and some asymmetry in rod length is introduced, the
  response of the metamolecule becomes a function
 of the relative rod lengths $H_s/H_l$.

 \subsection{Excitations of the toroidal dipole mode}
 \label{subsec:Amp}

 The eigenmodes in \fref{fig:Eigenmodes} are those of a symmetric
 toroidal metamolecule. When analyzing the amplitudes of the eigenmodes,
 of an asymmetric metamolecule, we do so
 using the symmetric metamolecule basis.
 The coupling matrix $\mathcal{C}$ is decomposed as
 \begin{equation}
 \mathcal{C}=\mathcal{C}_{\text{sym}}+\mathcal{A}
  \,\textrm{,}
 \label{eq:C_sym+A}
 \end{equation}
 where $\mathcal{C}_{\text{sym}}$ is the
 coupling between rods whose lengths are
 the mean rod length $H_0$. The matrix
 $\mathcal{A}$ contains the detail on asymmetry.
 The variation of the resonance frequencies
 between the rods generally suppresses the light-mediated
 interactions in the metamolecule\cite{PhysRevB.86.205128}.
 For the point dipole model, we define this
 as a diagonal matrix whose elements are
 the resonance frequency shifts of the different
 nanorods, see \aref{sec:asymm-coupl-coll}.
 In the finite-size resonator model, in addition to the
 resonance frequency shifts in the
 diagonals, the off-diagonal elements give the difference in the
 finite-size resonator interactions of a symmetric
 system and an asymmetric system, see \aref{sec:asymm-coupl-coll}.

 The amplitude of the different modes may be
 analyzed by expanding the vector of dynamic
 variables ${\bf b}(t)$ as
 \begin{equation}
  {\bf b}(t) = \sum_nc_n(t){\bf v}_n
  \,\textrm{,}
 \label{eq:c_iv_i}
 \end{equation}
 where $c_n(t)$ is the amplitude of the
 eigenmode ${\bf v}_n$ of $\mathcal{C}_{\text{sym}}$.
 We denote the toroidal dipole amplitude
 as $c_\text{t}$. In the absence of any
 asymmetry, the only modes driven by linearly
 polarized light are the E1 modes. When the
 asymmetry depicted in \fref{Drg:Asymmetry}
 is introduced to the metamolecule,
 the toroidal dipole mode (in addition to the E1 modes) is also driven.

 \begin{figure}[h!]
   \centering
 \includegraphics[width=0.9\columnwidth]
 {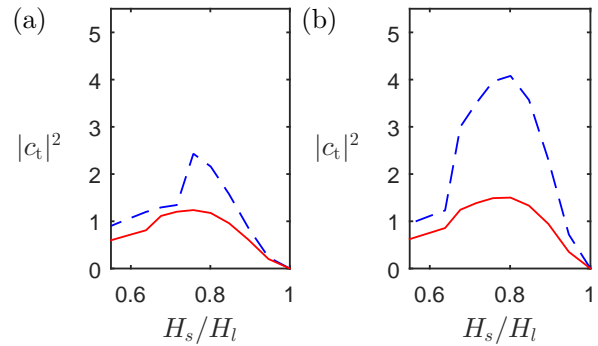}
 \caption{\label{fig:MaxAmp}(Color online)  The intensity of the collective
  toroidal dipole excitation as a function of the ratio of rod lengths $H_s/H_l$.
  The center of mass of the longer rods is at $s_l=\lambda_0/3$ and
  the layer position is
   $y = \lambda_0/10$.
   We show the point dipole model (blue dashed  lines)
   and finite-size model (red solid lines). The nanorod mean
   length and radius are those of the reference nanorod.
   The radiative emission rate in~(a) is
  $\Gamma_{\text{E}}=0.83\Gamma$. In~(b)
  there are no ohmic losses.
 }
 \end{figure}

 In \fref{fig:MaxAmp}, we show the maximum intensity
 of the toroidal dipole mode as a function of the asymmetry between
 the cylindrical rods, for the point
 dipole approximation and for the finite-size model,
 when the metamolecule is driven at the resonance of the toroidal mode
 of the symmetric metamolecule. If there are no ohmic losses, the finite-size
 model shows a maximum intensity when $H_s/H_l\approx 0.8$, and the
 intensity here of the point dipole model is
 approximately four times that of the
 finite-size model. As the asymmetry between the rods increases, the intensity
  of both the finite-size model and the point dipole approximation
 decreases.

 When ohmic losses are accounted for, the maximum intensity is
 when $H_s/H_l\approx 0.75$ before  decreasing.
 The incorporation of losses significantly affects the
 point dipole approximation, the maximum
  intensity is approximately $50\%$ less than when no losses were present.
  Conversely, the effect
 on the finite-size model is negligible.

 \begin{figure}[h!]
   \centering
 \includegraphics[width=0.9\columnwidth]
 {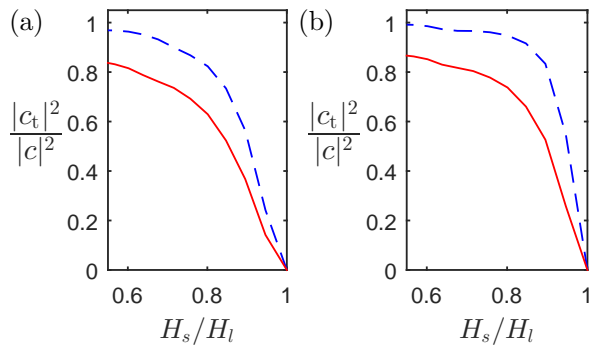}
 \caption{\label{fig:RelAmp}(Color online)  The  relative amplitude of
  the collective toroidal dipole excitation as a
 function of the ratio of rod lengths
  when driven on the toroidal dipole resonance.
The parameters as in Fig.~\ref{fig:MaxAmp}
 }
 \end{figure}

 In \fref{fig:RelAmp} the relative occupation of
 the toroidal dipole mode is shown as
 a function of the ratio of rod lengths.
 Although we consider a non unitary, open system,
 the eigenmodes have periodic boundary conditions
 and in the studied cases form a well-behaving
 orthonormal pseudo-basis. We define
 the overlap between an eigenmode $\colvec{v}_j$ with an
 excitation $\colvec{b}$ by
 \begin{equation}
   O_{j}({\bf b}) \equiv \frac{ | {\bf v}_j^T \colvec{b} |^2}
   {\sum_i | {\bf v}_i^T {\bf b} |^2} \label{eq:overlapDef}
  \,\textrm{,}
 \end{equation}
 where the summation runs over all the eigenmodes.
 In both cases, when losses are present
 and when they are neglected, the relative occupation
  of the point dipole approximation
  over-estimates the finite-size model. In the absence of ohmic losses, the
  relative occupation of the point dipole
 model saturates at $H_s/H_l\approx 0.8$,
 where the total excitation of the metamolecule is in the toroidal dipole mode.
 In the finite-size model when $\Gamma_{\text{O}}=0$, saturation occurs
 when $H_s/H_l\approx 0.7$ and the relative
 occupation is approximately $0.8$.
 When losses are present, the relative occupation
 at the maximum intensity of the
 toroidal dipole excitation ($H_s/H_l\approx 0.75$) is $0.95$ for the
 point dipole approximation. The finite-size model
 here shows a relative occupation of $0.75$.

  In the absence of asymmetry between the
 rods $H_s/H_l=1$, both the intensity
 plots in \fref{fig:MaxAmp}, and the relative occupation plots
 \fref{fig:RelAmp}, show that there is no toroidal dipole excitation.
 However, even a small asymmetry in rod
 lengths produces a toroidal dipole excitation.
  Although the intensity of the toroidal mode excitation
 can be maximized at a relatively small value of the rod
  asymmetry, the fidelity of the toroidal
 dipole mode keeps increasing when the asymmetry is increased.

 \subsection{Scattered light intensity in the far field}
 \label{subsec:scattered}

 \begin{figure}[h!]
 \centering
 \includegraphics[width=0.9\columnwidth]{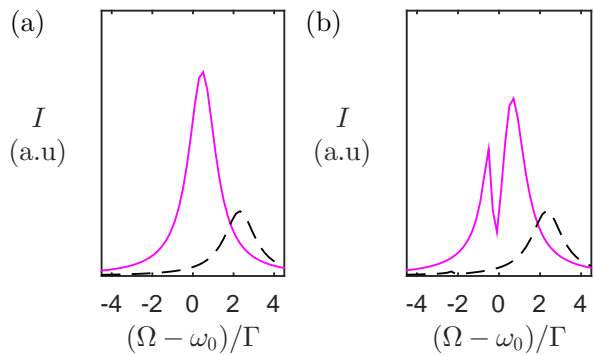}
 \caption{\label{fig:Intensity}(Color online)
 The scattered light intensity in the forward direction $I$, as a function of the
 detuning of the incident light from the resonance frequency
 $\omega_0$ of our reference nanorod.
 We show the responses of a symmetric
 toroidal metamolecule~(a) and a toroidal metamolecule with asymmetry
 promoting a toroidal response~(b). The radial position of the nanorods
 $s=\lambda_0/3$, the layer position $y=\lambda_0/20$ (black dashed lines)
 and $y=\lambda_0/10$ (magenta solid line), $H_s/H_l = 0.75$, and
 $\Gamma_{\text{E}}=0.83\Gamma$.  The
 intensity dip in (b) indicates a Fano resonance
 due to the interference between the E1 and toroidal dipole modes.
 The calculations are performed in the point dipole approximation.
 }
 \end{figure}

 It is also interesting to study the scattered light from a
 toroidal metamolecule in the far-field response.
 We again assume that the incident field propagating
 normal to the plane of the metamolecule excites the current
 oscillations in the nanorods. We calculate the collective
 excitations of the metamolecule by including all the
 radiative interactions between the nanorods.
 In \fref{fig:Intensity} we show the intensity of
 the scattered light in the forward direction
 (without the incident field contribution) from a
 symmetric toroidal metamolecule
 and from a toroidal metamolecule with the
 asymmetry designed to promote the toroidal
 dipole response. Two cases are displayed corresponding
 to two layer separations $y$.
 In the symmetric case, only  the E1 collective modes
 are excited, displaying
 broad resonances.

 For large $y$, the E1 resonance is close to the resonance
 frequency, $\omega_0$ of our reference nanorod,
 for small $y$ the resonance is blue-shifted.
 In the asymmetric case, the light excites the E1 modes and the toroidal
 dipole mode.
 A destructive interference between the broad-resonance
 E1 and the narrow-resonance toroidal modes produces
  a Fano resonance~\cite{fanoreview}.  For the case of a
  large layer separation $y$, the Fano resonance clearly
  shows up as a dip in the spectrum of the scattered
  light intensity at the resonance of the toroidal dipole mode,
 indicating suppressed forward scattering. This is because a
 toroidal dipole excitation on a plane normal to the propagation
 direction does not contribute to the far-field radiation.

 The interference of the Fano resonance has
 an analogy in atomic physics in the interference
of bright and dark modes in the electromagnetically-induced
 transparency\cite{FleischhauerEtAlRMP2005}.
 Here the subradiant toroidal mode (t) acts as a dark
 radiative mode and the superradiant electric dipole
 mode (E1) as a bright mode; in the excitation of the E1
 mode, the different scattering paths, $\rightarrow$E1,
 $\rightarrow$E1$\rightarrow$t$\rightarrow$E1, etc.,
 destructively interfere at the Fano resonance peak.
 The Fano resonances may also appear in other complex
  metamolecules, such as in oligomers\cite{GiessenOligomers},
   or as a result of a collective behavior of the metamaterial
   array\cite{PhysRevLett.111.147401}. The existence of more
   than one subradiant mode in a toroidal metamolecule and
 the possibility to employ collective effects in ensembles of
  toroidal metamolecules is particularly promising for tunable
  control of the resonances\cite{PhysRevLett.111.147401}
 and for sensing applications.

 Our simple model of the radiative intensity provides a
 qualitative description of the Fano resonance of the toroidal
 dipole mode in the forward-scattered far-field spectrum.
 For 2D metamaterial arrays of asymmetric split-ring
 metamolecules, the point dipole radiation model
 provides a good qualitative agreement with the
 experimental findings due to weak higher-order
 multipole radiation of individual split-ring
 arcs~\cite{JenkinsLineWidthNJP}. In the studied cases of the resonances of individual nanorods, higher-order
 multipole radiation is similarly weak. Comparisons between multipole expansions and complete field calculations were performed in \Rref{PhysRevB.89.205112} between the far-field radiation patterns of toroidal dipole resonances in a noninteracting resonator system.

 \section{Conclusions}\label{sec:Conclusions}

 We theoretically studied light-induced interactions in a toroidal
 metamolecule that consists of closely-spaced,
 strongly-coupled plasmonic nanorods.
 The interactions lead to collective excitation eigenmodes
 that exhibit collective resonance
 frequencies, linewidths, and line shifts. When the
 nanorod pairs are pointing radially outwards,
 one of the collective eigenmodes is identified as a toroidal dipole mode.
 We provided simple criteria to optimize a structural
 asymmetry of the metamolecule that allows a strong
 excitation of the toroidal dipole mode
 by a simple linearly polarized light beam.
 By analyzing a specific eight-rod case,
 we have shown how even small asymmetries lead
 to a large proportion of the total
 excitation to be found in the toroidal dipole
 mode.

 By comparing the point dipole approximation to a finite-size
  resonator model, we have shown that the point dipole approximation
 is sufficient to model interacting rods for large interrod separations,
 providing accurate descriptions when the layer and radial separations satisfy
 $y\agt\lambda_0/6$ and $s\agt \lambda_0/4$. For more closely-spaced rods,
 the nanorods' finite length and thickness  become increasingly important.

\begin{acknowledgments}
We acknowledge financial support from the EPSRC, the Leverhulme Trust, the Royal Society, and
the MOE Singapore Grant No. MOE2011-T3-1-005.
\end{acknowledgments}

 \appendix

 \section{Drude model}
 \label{sec:Drude}

 In this appendix,
 we consider the scattering and polarizability of small
 metallic nanorods in order to estimate
 the resonance frequency, as well as the radiative and
  ohmic decay rates of a single nanorod.
 We begin by considering the Drude model for the
  permittivity $\epsilon$ of a metallic rod\cite{AbsSca},
 \begin{equation}
 \epsilon(\omega)=\epsilon_\infty -
 \frac{\omega_p^2}{\omega(\omega + i\Gamma_\text{D})}
  \,\textrm{,}
 \label{app:eq:DrudeModel}
 \end{equation}
 where $\epsilon_\infty$ is the permittivity at
 infinite frequencies, $\omega_p$ is the
 plasma frequency and $\Gamma_\text{D}$ is the
 decay rate of current oscillations within
 the material.
 The scattering cross section of a small particle
 is dependent upon its polarizability
 $\alpha$\cite{AbsSca} and the wavelength $\lambda$ of the incident field
 \begin{equation}
 \sigma_\text{sc}=\frac{8\pi^3}{3\lambda^4}|\alpha|^2
  \,\textrm{.}
 \label{app:eq:sc_book}
 \end{equation}
 The polarizability depends on the physical
 characteristics of the particle including its
 volume $V_{\text{0}}$ and geometry, which is
 introduced through the depolarization
 factor $L$\cite{AbsSca}. In the Rayleigh approximation, the polarizability  is
 \begin{equation}
 \alpha_i=V_{\text{0}}\frac{\epsilon - 1}{1 + L_i(\epsilon - 1)},
 \qquad i = x,y,z
  \,\textrm{.}
 \label{app:eq:PolarizabilityRayleigh}
 \end{equation}
 The depolarization factor for a cylinder aligned along the $z$ axis
  is\cite{VIEmelyanov,Emelyanov2}
  \begin{equation}
 L_x=L_y=\frac{1}{2\sqrt{1 + \kappa^2}}
 \quad
 \text{and}
 \quad L_z = 1 - \frac{1}{\sqrt{1+ \kappa^2}}
  \,\textrm{,}
 \label{eq:L_i}
 \end{equation}
  where $\kappa=2a/H$ is the aspect ratio of the cylinder.
 The curve produced by the scattering cross section
 \eref{app:eq:sc_book} has two
 Lorentzian profiles with two independent resonance frequencies.
 There is a resonance representing
 the  longitudinal polarizability $\alpha_z$ with
 depolarization factor $L_z$, and a separate
  resonance for the radial polarizability
  $\alpha_x=\alpha_y$ with depolarization
  factors $L_x=L_y$.

  \begin{figure}[h!]
  \centering
 \includegraphics[width=0.9\columnwidth]
 {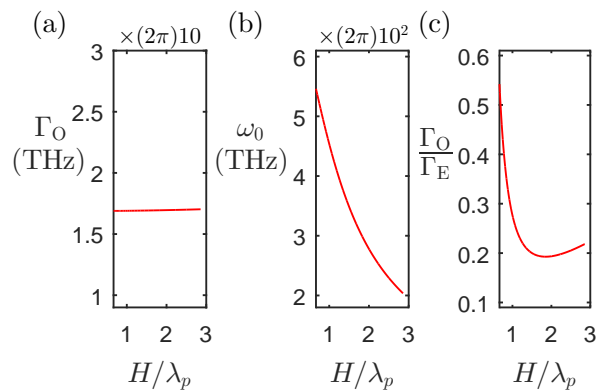}
 \caption{\label{fig:GammaO_Omega0_RelativeGamma_E1}(Color online)
 The ohmic losses,
 resonance frequency, and relative radiative decay rate
 as a function of the rod length for a gold nanorod with radius
 $a=\lambda_p/5$. We show the ohmic
 losses $\Gamma_\text{O}$ in~(a),  the resonance
 frequency  $\omega_0$ in~(b) and the relative radiative decay
 rate $\Gamma_\text{O}/\Gamma_\text{E}$ in~(c).
 }
 \end{figure}

   For gold, the Drude parameters are:   \cite{PhysRevB.6.4370Christy,doi:10.1021/j100287a028Schatz,Grady2004167}
   $\epsilon_\infty = 9.5$;  $\omega_p = (2\pi)2200$~THz;
   and $\Gamma_\text{D} = (2\pi)17$~THz.  In the Rayleigh
   approximation the full width at
   half maximum (FWHM) of the scattering cross section
  \eref{app:eq:sc_book} is approximately
 independent of the length of the rod and
  gives the value of the ohmic loss rate $\Gamma_\text{O}$.
  In  \fsref{fig:GammaO_Omega0_RelativeGamma_E1}(a)
  and~\ref{fig:GammaO_Omega0_RelativeGamma_E1}(b),
 we show $\Gamma_\text{O}$
   and resonance frequency $\omega_0$ for
  a gold nanorod with radius $a=\lambda_p/5$,
  where $\lambda_p = 2\pi c/\omega_p\simeq 139$nm.
 We find $
  \Gamma_\text{O} \simeq \Gamma_\text{D}\simeq (2\pi)17~\text{THz}$.

 For larger particles, the Rayleigh approximation is
 insufficient and retardation effects must be considered.
  Mie's formulation accounts for this
 retardation for spherical particles. In \Rref{kuwate},
 a generalization of Mie's
 polarizability is obtained for nonspheroidal particles
 that has been used successfully to
 model the scattering of metallic nanoparticles.
 The approximate ratio of ohmic losses to the radiative decay rate is
 \begin{eqnarray}
 \frac{\Gamma_\text{O}}{\Gamma_\text{E}} &=&
  -\text{Im}
 \left(\dfrac{3\lambda_0^3}{4\pi^3a^2H(\epsilon - 1)}\right)
 \nonumber\\
 &=&\frac{6\Gamma_\text{D}\omega_p^2c^3}
 {a^2H\omega_0^2
 \left[\omega_0^2\Gamma_\text{D}^2
 (\epsilon_\infty - 1)
 +
 \left(\omega_0^2(\epsilon_\infty - 1) - \omega_p^2\right)^2
 \right]}
  \,\textrm{.}\nonumber\\
 \end{eqnarray}
 In \fref{fig:GammaO_Omega0_RelativeGamma_E1}(c),
 we show the relative decay rates as a function of the rod length. For shorter
 rods $H<\lambda_p$, the ohmic losses
   are dominant. For longer rods the radiative emission rate
 is dominant.
   For $\lambda_p<H<3\lambda_p$ the radiative decay rate is
  approximately constant, $
  \Gamma_{\text{E}}\approx 5\Gamma_\text{O}$.
  In \sref{sec:T1}, we model the interactions between nanorods
  as point dipoles and as finite-size cylinders using this value.
     In \ssref{subsec:Driving}
  and~\ref{subsec:Amp}, where the length of the nanorod is important,
  the resonance frequency of each nanorod is
  assumed to depend on the length of the rod as
  shown in \fref{fig:GammaO_Omega0_RelativeGamma_E1}.

 \section{Asymmetric Coupling of Collective Modes}
 \label{sec:asymm-coupl-coll}

 In \sref{subsec:Amp}, we showed how linearly polarized light could drive a
  toroidal dipole response in a metamolecule whose constituent rods
 vary in resonance frequency.
 We found an optimal variation in the limit that the
 resonance frequency shifts were much larger than the rod linewidths,
 and interactions between individual rods can be neglected.
 In this appendix, we show how this scheme also works in the presence of
 interactions.
 Describing the evolution of the system in terms of collective
 eigenmodes of a symmetric metamolecule, we see how the introduced
 asymmetry couples collective metamolecule modes.
 We will see that a judicious combination of rod lengths can strongly
 couple an electric dipole mode (driven by the incident
 linearly polarized light) to the toroidal dipole mode (which is invisible to the
 incident field without rod asymmetries).

 Formally, the evolution of the resonator excitations can be expressed
 in terms of the driven, system of equations \eref{eq:equmot}.
 When all rods are of equal length $H_{j}=H_{0}$, the coupling matrix
 $\mathcal{C}=\mathcal{C}_{\mathrm{sym}}$, and we denote the driving of
 the resonators as $\colvec{F}_{\mathrm{sym}}$, with elements
 $F_{\mathrm{sym},j}$ given in \eref{eq:f_in}.
 As discussed in \sref{sec:symm-toro-metam}, this symmetric
 metamolecule has eigenmodes of oscillation (labelled by index $n$)
 corresponding to eigenvectors ${\colvec{v}}_{n}$
 of $\mathcal{C}_\text{sym}$
 and eigenvalues $\xi_{n}=-i\delta_{n}-\gamma_{n}/2$, where
 $\delta_{n}$ is the shift of the collective mode resonance frequency
 from the reference $\Omega_0$,  and $\gamma_{n}$ is the collective
 decay rate, \eref{eq:lambda_n}.
 Generally, any metamolecule excitation $b$ can be expressed as
 \begin{equation}
   {\colvec{b}}   =\sum_{n}c_{n}(t) {\colvec{v}}_{n}
                = \mathcal{S} {\colvec{c}}
               \, \textrm{,}
  \label{eq:app_eigen_expansion}
 \end{equation}
 where $\mathcal{S}$ is a matrix whose $n$th column is the eigenvector
 ${\colvec{v}}_{n}$ of $\mathcal{C}_{\mathrm{sym}}$, and ${\colvec{c}}
 \equiv (c_1, \ldots, c_n)^T$.
 The electric dipole $\colvec{v}_{\text{E1}}$ and
 toroidal dipole $\colvec{v}_{\text{t}}$ excitations
 are eigenmodes of the symmetric metamolecule,
 where $\colvec{v}_{\text{E1}}$ is directly driven by the incident field.
 When the two layers of rods are separated by much less than a
 wavelength, the incident field drives only one of the two modes
 where the adjacent dipoles oscillate symmetrically,
 leaving all of the other modes, including the toroidal
 dipole unexcited.

 Here we generalize the treatment of \sref{subsec:Driving}, which
 dealt with the limit of non-interacting rods, to show how introducing
 an asymmetry into the rod lengths can lead to the excitation of a
 toroidal dipole.
 Perturbing the lengths of each rod $j$ by $\delta H_{j}$ during the
 fabrication, alters the coupling matrix $\mathcal{C}$.
 As discussed in \sref{subsec:Driving}, the primary consequence of
 altering the rod lengths is that each rod has its resonance frequency
 shifted by $\delta\omega_{j}$ proportional to $\delta H_j$,
 as indicated in \fref{fig:GammaO_Omega0_RelativeGamma_E1}.
 Additionally, changing rod lengths impacts the interactions between
  metamolecules.
 We denote the deviation of the coupling matrix from that for the
 symmetric system as
 \begin{equation}
   \mathcal{A} \equiv \mathcal{C} - \mathcal{C}_{\mathrm{sym}}
  \,  \textrm{.}
  \label{eq:AppAsymMatDef}
 \end{equation}
 From \eref{eq:equmot}, the amplitude of each amplitude $c_{n}$
 in the expansion of \eref{eq:app_eigen_expansion} obeys
 \begin{equation}
   \dot{\colvec{c}}=\left[ \Lambda + \mathcal{S}^{-1}
     \mathcal{A}\mathcal{S}\right]{\colvec{c} }+ {\colvec{f}} +
   \colvec{\delta f}
   \, \textrm{,}
 \end{equation}
 where $\colvec{f} \equiv \mathcal{S}^{-1}
 \colvec{F}_{\mathrm{sym}}$ is the vector of
 driving amplitudes for each individual mode, and $\Lambda$ is the
 diagonal matrix of eigenvectors of $\mathcal{C}_{\mathrm{sym}}$.
 Changing the rod lengths also alters the driving by $\colvec{\delta f}
 \equiv \mathcal{S}^{-1} \colvec{\delta F}$, where $\delta F_j = \delta
 F_j \cos\theta_j$ is the change in driving amplitude experienced by
 rod $j$, and $\delta F_j$ is the change in driving amplitude a rod
 would experience if it were oriented along the incident field
 polarization, as discussed in \sref{subsec:Driving}.
 Essentially, altering the lengths of the rods induces coupling between
 the eigenmodes via the asymmetry matrix $\mathcal{S}^{-1}
 \mathcal{A}\mathcal{S}$.  At the same time, whilst a linearly polarized
 incident field only drives electric dipole modes in the symmetric
 metamolecule, a non-zero $\colvec{\delta f}$ introduced by the
 asymmetry permits other modes to be driven directly by the incident
 field.

 As in \sref{subsec:Driving}, our goal in altering the lengths is
 to find a perturbation that permits the excitation of the toroidal
 dipole mode, while reducing the contribution of other collective
 modes.
 In particular, consider the steady-state excitation induced by a field
 resonant on the toroidal dipole of the symmetric system
 \[
 {\colvec{c}}=-\left[  \Lambda+i\delta_{\text{t}} +
   \mathcal{S}^{-1}\mathcal{A}\mathcal{S}
 \right]^{-1} \left(\colvec{f} + \colvec{\delta f}\right)
 \, \textrm{,}
 \]
 where the subscript ``t" refers to the toroidal dipole mode.
 Since altering the lengths of the rods induces coupling between the
 modes, one can produce a toroidal dipole excitation by introducing a
 coupling between the toroidal dipole mode and other modes in the
 metamolecule,  in particular the  electric dipole mode.
 In general, one would obtain the optimal excitation of the toroidal
 dipole by optimizing the rod length perturbations $\delta H_{j}$.
 The general optimization procedure would account for changes in
 interactions between resonators produced by the asymmetry
 as well as the interactions that
 are present in the symmetric metamolecule.

 Here, we illustrate how rod lengths would be chosen when the only effect of
 $\delta H_{j}$ is to produce changes in resonance frequencies $\delta
 \omega_j = \chi \delta H_j$ and individual rod decay rates
 $\delta\Gamma_j =  \nu\delta H_j$ in
 proportion to $\delta H_j$ for some constants $\chi$ and $\nu$.
 In this case, the matrix $\mathcal{A}=\operatorname{diag}\left(
   -\delta\omega_{1} - \delta\Gamma_1/2, \ldots, -i\delta\omega_{N} -
   \delta\Gamma_N/2\right)$.
 We will also assume the separation between rod layers is much less
 than a wavelength so that each layer experiences an identical
 driving.
 In these limits, we ask the question: what conditions would have to be
 satisfied to have the steady-state response of the metamolecule to be
 purely in the toroidal dipole mode?
 From there, we deduce a combination of $\delta H_{j}$, that could
 yield a toroidal dipole
 response.
 Consider an excitation of the form ${\colvec{b}} = c_{0}
 {\colvec{v}}_{\text{t}}$ at time $t=t_{0}$, entirely in the toroidal dipole
 mode.
 Then, from \esref{eq:equmot},~\eqref{eq:f_in}, and~\eqref{eq:F_tot_j},
 we have, for each resonator $j$,
 \begin{eqnarray}
   \left. \frac{d}{dt} b_{j}\right\vert _{t=t_{0}} &=&
   - c_{0}\frac{\gamma_{\text{t}}}{2} {\colvec{v}}_{\text{t}}(j) - i
   c_{0} (\chi - i \nu)\delta H_{j}
   {\colvec{v}}_{\text{t}}(j) \nonumber \\
   & &+ (F_0 + \delta F_j)\cos \theta_j
   \,  \textrm{.}
   \label{eq:app_steady_state}
 \end{eqnarray}
 From \eref{eq:app_steady_state}, we see
 that if $\gamma_\text{t}$ is negligible, i.e.,
 $\gamma_{\text{t}} \ll \delta\omega_{j}$,
 $\Gamma$, the toroidal excitation
 is in the steady state when for each nanorod $j$, $\delta H_j$
 solves
 \begin{equation}
   \label{eq:solution_general_ish}
   i  c_{0} (\chi - i \nu)\delta H_{j}
   {\colvec{v}}_{\text{t}}(j) = (F_0 + \delta F_j)\cos \theta_j
    \,\textrm{.}
 \end{equation}
 As in \sref{subsec:Driving}, where we considered non-interacting
 metamolecules, to lowest order in $\delta H_j$, the asymmetry in rod
 lengths needed to generate a toroidal dipole is given by
 \eref{eq:general_rod_length_profile}.

 Thus, when the collective decay rate
 $\gamma_{\mathrm{t}} \ll \delta\omega_j$, and we have neglected how
 the asymmetry of rod lengths alters interactions between the rods, the
 asymmetry of \eref{eq:general_rod_length_profile}
 would yield a toroidal dipole
 amplitude
 \[
  c _0 = -i \sqrt{N} \frac{F_{0}} {(\chi - i \nu)\delta H_{0}} \,\textrm{.}
 \]
 This is remarkably similar in form to the toroidal dipole amplitude one would
 obtain if one neglected all interactions between nanorods as done in
 \sref{subsec:Driving}.
 In accounting for interactions, we no longer need to assume that
 $\delta\omega_j \gg \Gamma$.  One does, however, need to drive the
 metamolecule with a field resonant on the toroidal dipole mode of the
 symmetric metamolecule, rather than resonant with a single nanorod.

\end{document}